# Metasurface-Assisted Adaptive Quantum Phase Contrast Imaging

Xiaojing Feng,[a,†] Juanzi He, [a,†] Xingyu Liu, [a,†] Xiaoshu Zhu,[a] Yifan Zhou,[a] Xinyang Feng,[a] Shuming Wang[a,*]

[a]National Laboratory of Solid-State Microstructures, Collaborative Innovation Center for Advanced Microstructures, School of Physics, Nanjing University, Nanjing, 210093, China

**Abstract**. Quantum imaging employs the nonclassical correlation of photons to break through the noise limitation of classical imaging, realizing high sensitivity, high SNR imaging and multifunctional image processing. To enhance the flexibility and imaging performance of the optical systems, metasurfaces composed of subwavelength structural units provide a powerful optimization approach, enabling advanced applications in quantum state modulation and high-precision imaging. Conventional phase contrast imaging is fundamentally constrained by its single-phase modulation scheme, precluding adaptive switching between imaging modalities. Therefore, the development of high-contrast imaging techniques that can be used in any combination of phases has been a challenge in the field of optical imaging. Here, we propose a novel imaging scheme combining a polarization-entangled light source and a polarization multiplexed metasurface, which realizes remotely switchable bright-dark phase contrast imaging, demonstrating the flexibility and high integration of the system. Experiments show the system can realize high contrast and high SNR imaging under low phase gradient conditions (phase difference as low as $\pi/5$) and exhibit excellent phase resolution. In addition, the system is suitable for imaging biological samples under low-throughput light conditions, providing an efficient and non-destructive shooting solution for biomedical imaging and promoting the development of phase-sensitive imaging technology.

*Shuming Wang**,** E-mail: wangshuming@nju.edu.cn

## 1  Introduction

Two key concepts that distinguish quantum optics from classical optics are quantum entanglement[1–4] and quantum interference [5–8] . Based on these two fundamentals, emerging quantum technologies such as quantum cryptography[9,10], invisible state transfer[11–13], super-resolution metrology[14] and quantum imaging[15–19] have gradually developed. Quantum imaging[20] is a cutting-edge technique for image acquisition by exploiting nonclassical correlations (e.g., entanglement and statistical correlation) between photons. This technique enables high-contrast imaging under low-light illumination and captures phase information[17,18,21], and high-dimensional optical parameters[19,22], that cannot be detected by conventional imaging methods, thus breaking through the noise limitations of classical imaging and achieving higher sensitivity, resolution, and signal-to-noise ratio[16,23,24]. However, existing quantum imaging systems usually face the problems of bulky size, complex design and single function, which limit their compactness and versatility in practical applications. Therefore, the development of highly integrated, multi-functional and adaptable quantum imaging systems has become an important research direction to promote the field towards practical application.

Recently, optical metasurfaces have received much attention due to their unique advantages of miniaturization and integration [25–27]. As two-dimensional material consisting of array of subwavelength structures integrated on a chip, the metasurface realize accurate modulation of multiple parameters such as phase, polarization and frequency of light waves through precisely designed arrangement of metatoms, demonstrating an excellent optical field modulation capability [28–31]. Currently, metasurfaces have been widely used in fields of metalens[32,33], optical achromats[34–36], optical cloaking devices[37,38] and topological photonic systems[39,40]. Based on this, the combination of quantum optics and metasurfaces can achieve high efficiency, miniaturization, and flexibility in optical imaging, offering new prospects for the development of both fields [22,26]. Previously, the multiparameter modulation capability of metasurfaces combined with multidimensional entangled photons[41–43], has led to significant progress in the fields of quantum light source[43,44], quantum state manipulation[45,46] and quantum sensing[47].

In the field of biomedical imaging, the observation of pure phase samples (e.g., tissue slices and living cells) has always been a challenge. Conventional brightfield microscopy relies heavily on differences in light absorption by the sample to generate images. However, many biological samples are nearly transparent in visible light, resulting in low imaging contrast, and conventional staining methods may not only affect cellular activity but also limit the ability to image live cells in real time [48,49]. Based on this, Zelnick first proposed the phase contrast imaging technique, which uses the interference effect of light to convert the phase distribution of a sample into an intensity distribution by introducing the phase difference between scattered light and transmitted light of the sample, which significantly enhances the contrast on imaging surface, thus realizing accurate measurement and visualization of purely in-phase objects[50]. However, conventional phase contrast imaging techniques are limited by single phase modulation mode, which can only obtain a fixed imaging contrast distribution. This technical bottleneck stems from the strong dependence of imaging contrast on phase modulation, where a specific degree of phase change leads to an inversion of the light-dark distribution on the imaging surface, and the response of different sample structures to the same phase modulation varies[51,52]. Therefore, the development of new phase contrast imaging methods with high phase adaptability to achieve active modulation of imaging contrast remains an important scientific issue in the field of modern optical microscopy that needs to be broken through.

In this study, we propose a novel imaging scheme that is based on a polarization entangled source with a polarization multiplexed metasurface. The proposed scheme has the ability to realize remotely switchable bright-dark phase contrast imaging functions. In addition, the scheme is able to achieve high-contrast and high signal-to-noise imaging of low phase gradient samples under arbitrary phase combinations. The proposed scheme successfully addresses the technological limitations of conventional phase contrast imaging techniques, particularly their restricted contrast variability and limited flexibility. By combining the polarization entanglement state with the phase modulation property of the metasurface, we transform the polarization entanglement between twin photons into a dynamically modulated correlation of bright and dark phase images in the signal arm, which enables flexible switching of imaging effects. In the experiment, we successfully realized remote switching of bright and dark phase contrast images in the signal arm by projecting the heralding photons to different polarization basis, demonstrating the flexibility and high integration of this imaging system. In the field of biomedicine, the system utilizes the entanglement characteristics and high signal-to-noise ratio of quantum entangled light source, combined with precise modulation capability of metasurface on the phase, amplitude and polarization of light, to break through the technological bottlenecks of traditional imaging methods in terms of a single imaging effect, low resolution and label-free detection. With high signal-to-noise imaging under low light conditions, the system provides an innovative solution for non-destructive observation of biological samples, and shows great potential for future research in areas of brain neural network activity, bio-imaging, and life sciences.

## 2    Principle of Quantum Bright-Dark Phase Contrast Imaging

We first briefly review the basic principles of bright-dark phase contrast imaging based on classical continuous wave(CW) illumination. Subsequently, the method for remotely switching image modes of the system following the utilization of polarization multiplexed metasurface and quantum polarization entanglement source is delineated.

*2.1 Bright-dark phase contrast imaging with metasurface*

Phase-contrast microscopy enables the visualization of transparent samples by converting subtle phase shifts induced by sample structures into measurable intensity variations. The technique exploits optical interference phenomena - when coherent illumination passes through a phase object, the resulting scattered and unscattered wavefronts acquire different phase delays. By introducing a controlled phase shift between these wavefronts using specialized optical components, their subsequent interference produces a detectable amplitude contrast. The phase-to-amplitude conversion mechanism resolves sub-wavelength structural information beyond the diffraction limit that would otherwise remain invisible in conventional bright-field microscopy. Importantly, the contrast can be optimized by precisely tuning the phase offset, allowing clear imaging of weakly scattering biological specimens without the need for staining or labelling.

The key to bright-dark phase contrast imaging is the introduction of phase modulation on the Fourier plane. Based on this, we place the metasurface on the Fourier plane of the pure phase samples, and the mode switching of bright-dark contrast imaging can be realized by using the polarization-multiplexing metasurface to introduce different phases to the incident light under different polarizations. The incident light field is $E_0$, and the light field becomes $E_0 e^{i\varphi(x,y)}$ after passing through a purely phase sample with phase distribution $\varphi(x, y)$. Placing the metasurface on the Fourier plane modulates the light field in different regions, i.e., modulates different parts of the frequency space information. Assuming that the part of the incident light field modulated by the metasurface on the Fourier plane is $E_m$, the phase difference introduced is $\phi_m$, and the unmodulated part accumulates phase on the Fourier plane is $\phi_n$, and the phase difference between the two parts is $\Delta\phi = \phi_m - \phi_n$. Then the phase samples are imaged by the 4f system, and the light intensity distribution on the surface can be expressed as:

$$\begin{aligned} I &= \left|\left(E_0 e^{i\varphi(x,y)} - E_m\right)e^{i\phi_n} + E_m e^{i\phi_m}\right|^2 = \left|\left(E_0 e^{i\varphi(x,y)} - E_m\right) + E_m e^{i\Delta\phi}\right|^2 \\ &= |E_0|^2 + 4\sin^2\frac{\Delta\phi}{2}|E_m|^2 + \left(e^{-i\Delta\phi} - 1\right)E_0 e^{i\varphi(x,y)} E^*_m + \left(e^{i\Delta\phi} - 1\right)E_0 e^{-i\varphi(x,y)} E_m \end{aligned} \quad (1)$$

If we make $E_0 e^{i\varphi(x,y)} E^*_m = |E_0||E_m|e^{i\alpha}$, where $\alpha = \varphi(x, y)$, is the function of x and y, then the light intensity distribution is expressed as:

$$\begin{aligned} I &= |E_0|^2 + 4\sin^2\frac{\Delta\phi}{2}|E_m|^2 + \left(e^{-i\Delta\phi} - 1\right)|E_0||E_m|e^{i\alpha} + \left(e^{i\Delta\phi} - 1\right)|E_0||E_m|e^{-i\alpha} \\ &= |E_0|^2 + 4\sin^2\frac{\Delta\phi}{2}|E_m|^2 + 2|E_0||E_m|[\cos(\Delta\phi - 1)\cos\alpha + \sin\Delta\phi\cos\alpha] \end{aligned} \quad (2)$$

Therefore, the phase information at different positions on a pure phase sample is the key factor leading to the ups and downs of imaging intensity. However, even if a phase difference is introduced but the phase value of the pattern is different, the imaging effect will not be the same because a phase difference cannot make a variety of phase patterns have a better contrast effect. When the phase information is the same, the incidence of light in varying polarization states engenders disparate phase difference $\Delta\phi$. Consequently, this gives rise to divergent imaging contrasts. When the phase difference change $\pi$, the original bright spot may become darker, changing from bright phase contrast to dark phase contrast, that is, switching between bright-dark phase contrast imaging modes. The so called bright-dark contrast control is actually the switching of the two phase differences. This is the rationale behind the switching of the two phase contrasts.

## 2.2 Principle of Quantum Remote Switching

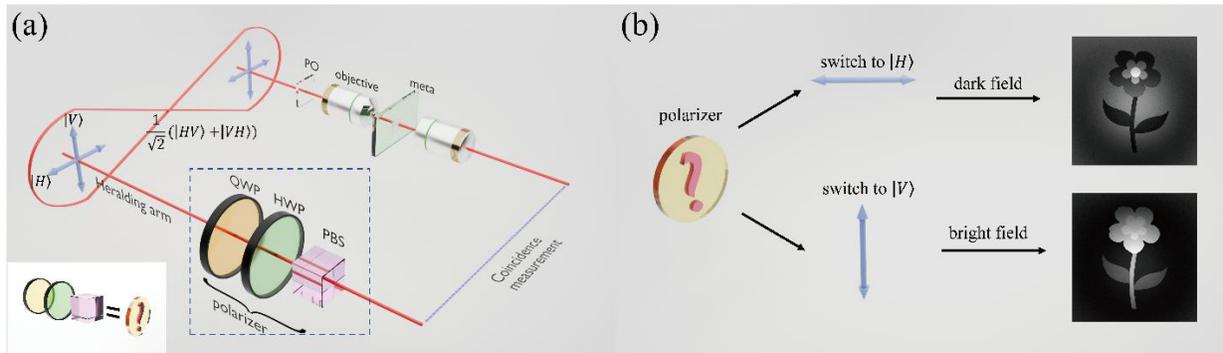

**Fig. 1 Schematic representation of a switchable quantum bright-dark phase contrast imaging system.** (a) Metasurface facilitates bright-dark phase contrast imaging capabilities. The quantum state of a polarization-entangled photon pair is $1/\sqrt{2}(|HV\rangle + |VH\rangle)$, where $|H\rangle(|V\rangle)$ represents the horizontal (vertical) polarization state. The phase sample and the metasurface are located in the imaging arm, and the signal photons are incident on the metasurface after passing through the phase object (PO) and are modulated to be imaged. The choice of polarization on the heralding arm is unknown at the moment, and the imaging arm outputs an image in the form of a quantum superposition of the bright-dark phase contrast images. (b) The heralding arm remotely manipulates the bright-dark phase contrast imaging modes. When the polarization basis vector of the heralded photon is projected to $|H\rangle$ polarization, the image acquired by the ICCD in the imaging arm is a dark image; when the polarization basis vector

of the heralding photon is projected to $|V\rangle$ polarization, the image is a bright image. The bright and dark images on the right are theoretical results obtained by simulation using MATLAB software.

We have successfully realized a remotely switchable quantum bright-dark phase contrast imaging scheme by utilizing polarization-entangled photon pairs. As shown in Fig. 1(a), the imaging system adopts a dual-optical-path design, including two functional modules, the imaging arm and the heralding arm. Among them, the polarization-entangled photons in the imaging arm are precisely modulated for sample illumination, while the herald arm realizes the remote switching control of the imaging mode through quantum state projection measurement. When the incident light from the imaging arm is $|H\rangle$, the center of light field through the metasurface is introduced an additional phase of $\pi/2$; when the incident light is in $|V\rangle$ polarization state, the introduced additional phase is $3\pi/2$. The light field modulated by the metasurface interferes with the unmodulated background light, forming a bright-dark difference phase contrast image, as shown in Fig. 1(b). In this study, we use a polarization-entangled state represented as $1/\sqrt{2}(|HV\rangle + |VH\rangle)$ as the illumination source, and when the input photon state in the heralding arm is set to $|H\rangle$ or $|V\rangle$, the input photon state in the signal arm will also be $|V\rangle$ or $|H\rangle$. This allows the heralding arm to act as a trigger for remotely switching that determines the bright or dark phase imaging mode of the signal arm.

## 2.3 Sample design and characterization

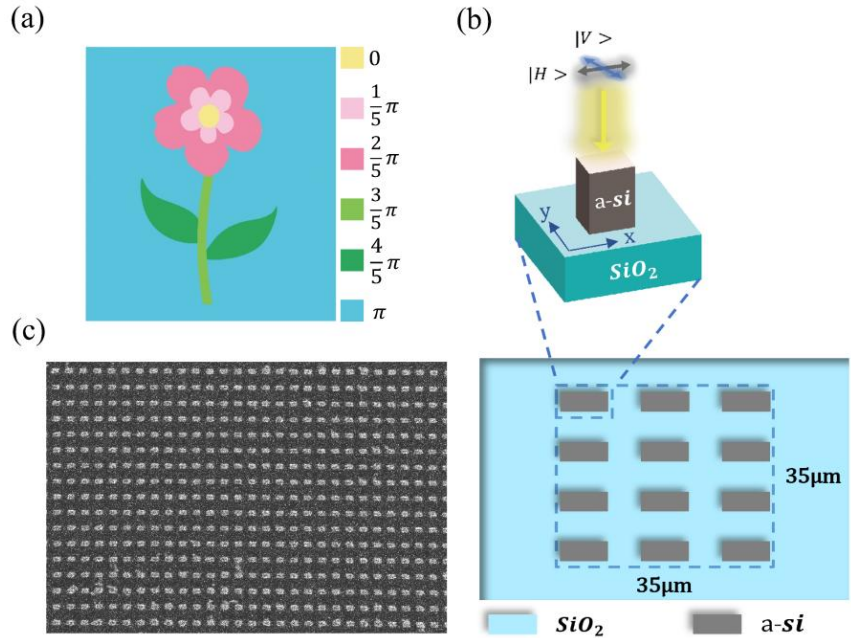

**Fig. 2 Characterization of experimental samples. (a)Low phase gradient samples.** The phase shifts of the different color coded regions are all π/5。The phase at the yellow petals is 0; the phase at the first petal layer (light pink) is π/5; the phase at the second petal layer (dark pink) is 2π/5; the phase at the flower stems (light green) is 3π/5; the phase at the flower leaves (dark green) is 4π/5 and the phase at the blue background is π. (b) Schematic diagram of the metasurface. The metasurface has a size of 35μm*35μm and a structural period of 350nm, exhibiting a periodic arrangement of silicon nitride nanopillars with dimensions of 190 nm in length, 110 nm in width, and 390 nm in height. (c) Top-view SEM image of the metasurface.

The low phase gradient sample used in this method is shown in Fig. 2(a), which features a flower with phase shifts in the color coded regions all differing by π/5，with phases of 0、π/5、2π/5 from the flower core to the outermost petals, and phases of 3π/5 和 4π/5 for the stems and leaves, respectively, and a blue background with a phase of π. As illustrated in Fig. 2(b), the unit structure of the metasurface is depicted schematically. The metasurface is supported by silicon dioxide, on which amorphous silicon pillars measuring 190 nm in length, 110 nm in width, and 390 nm in height are arranged at a period of 350 nm, with an overall size of 35 µm*35 µm. Fig. 2(c) presents a scanning electron microscope (SEM) image of the metasurface. We use commercial software Finite Difference Time Domain (FDTD) to compute the variation of the outgoing light after a plane wave is incident on the unit structure to obtain its modulation in phase and amplitude, applying periodic boundary conditions along the x and y directions and setting up a perfectly matched layer (PML) along the z direction, and scanning the length and width of the column to obtain the additional phase and transmittance for the horizontal and vertical directions of the column. Based on this, we can find a set of suitable sizes of the unit structure to be arranged to obtain the polarization-multiplexing metasurface we need, so that different phase modulations can be obtained after the incident of different polarized light. The size of the metasurface and the size of the phase sample directly determine the phase contrast imaging effect. After simulating different sizes of the metasurface and the phase sample in MATLAB software, the size of the metasurface is finally determined to be 35µm and the size of the phase sample is determined to be 320µm, which achieves the optimal bright-dark phase contrast imaging effect.

# 3 Results

## 3.1 Experimental setup

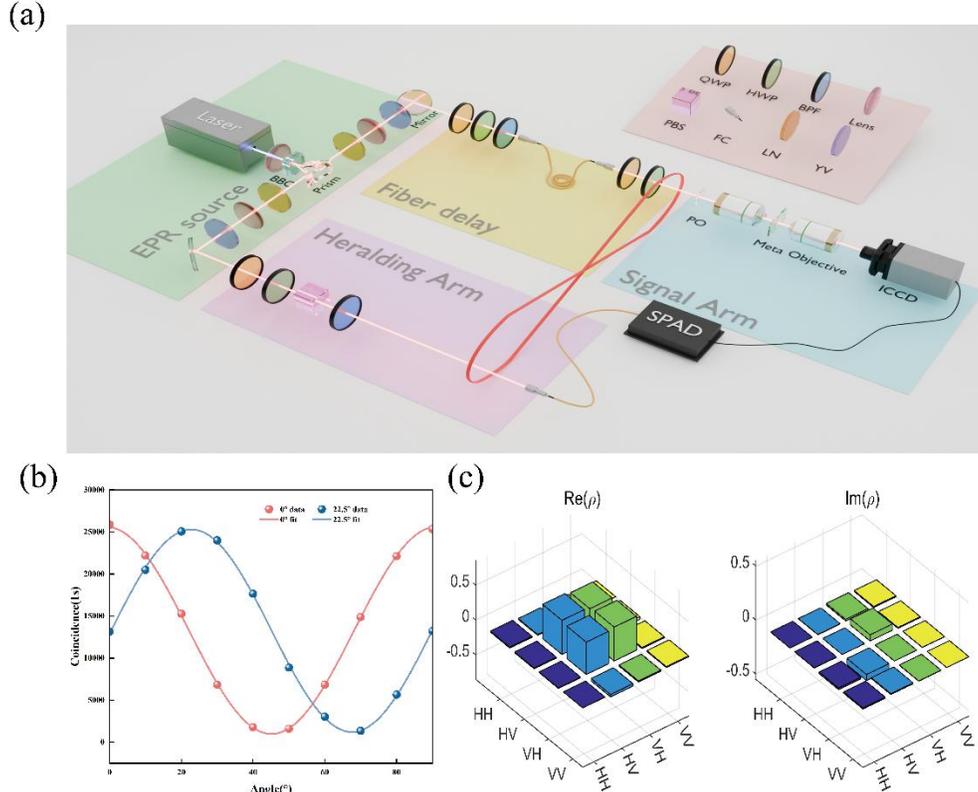

**Fig. 3 Experimental setup and characterization of entangled light source.** (a) Experimental setup. A 390 nm pulsed ultraviolet laser was passed through a lens and two BBO crystals arranged in a sandwich structure to generate pairs of entangled photons by spontaneous parametric down-conversion. Each pair of entangled photons is indistinguishable after being through the time-compensated and space-compensated crystal. The QHP is used to examine and select the polarization state of the heralding photons, which subsequently reach the SPAD trigger ICCD; the signal photons is phase-sampled and then modulated by the metasurface, then finally collected for imaging by the ICCD. BBO: Barium beta-borate crystals; HWP: half-wave plate; QWP: quarter-wave plate; FC: fiber coupler; PBS: Polarizing beam splitter; BPF: bandpass filter; YV: Yttrium Vanadate Crystals ($YVO_4$) for space compensation; LN：Lithium niobate crystals ($LiNbO_3$) for time compensation; PO: phase object. (b) Sinusoidal fit of the data when the HWP in the signal arm is fixed at 0° (red) and 22.5° (blue). (c) Real and imaginary parts of the density matrix of the quantum state.

The experimental setup is shown in Fig.3(a). The pump source is a high intensity pulsed laser that produces a focused 390 nm femtosecond laser beam with very high instantaneous power for efficient excitation of the SPDC. The pump light is focused by a convex lens into a 'sandwich' structure of Type II BBO crystals, e.g., a 1/2 wave plate sandwiched between two barium metaborate crystals, and the pump light generates a pair of parametric photons with the photon states of $|H\rangle_1|V\rangle_2$ by spontaneous parametric down-conversion in the first layer of the crystals, and then the photon states of the first pair of parametric photons are transformed to $|V\rangle_1|H\rangle_2$ after passing through the 1/2 wave plate, followed by another pump light in the second layer of the crystals. The pump light then excites another pair of parametric photons in the $|H\rangle_3|V\rangle_4$ photon state in the second crystal layer, and both pairs of parametric photons have a wavelength of 780 nm. Each pair of parametric photons is reflected by the triangular prism to the heralding arm (left) and the signal arm (right), respectively. Since the two photons divided into the same arm come from different pairs of parametric photons, they are generated at different temporal and spatial positions. In addition, different polarized light propagates in different directions and at different speeds in spatially anisotropic crystals. Therefore, the Lithium niobate crystal ($LiNbO_3$) and the Yttrium Vanadate Crystals ($YVO_4$) are placed in each of the heralding and signal arms to compensate for the temporal and spatial walk-off of the parametric light to achieve the indistinguishability of the photons. Simultaneously, a lens positioned between the crystals is used to collimate the optical path. At this point, the two-photon state interferes, resulting in the final output parametric light state $1/\sqrt{2}(|H\rangle_i|V\rangle_s + |V\rangle_i|H\rangle_s)$, where the subscripts i and s denote the position of the photon in the heralding and signal arms, respectively.

A long-pass filter and a narrow-band filter are placed in both the heralding and signal arms to eliminate undesirable fundamental frequency light. In the heralding arm, the heralding photons first pass through a QHP combination (i.e., a combination of a 1/4 wave plate and a 1/2 wave plate), and then pass through a polarized beam splitter (PBS). The PBS selects the photon polarization and projects the polarization state onto a specific polarization base, thus switching between bright-field imaging and dark-field imaging. Subsequently, a single photon avalanche detector (SPAD) is employed to detect the optical signal, thereby converting it into an electrical signal. This triggers the DDG mode of the enhanced charge-coupled device (CCD) camera to capture twin signal photons for imaging. Since there is some time delay in this process, to ensure that the signal photons captured by the ICCD and the heralding photons received by the SPAD are from the same pair, we use a 20m single-mode optical fiber in the optical path of the signal arm, and set up a nanosecond electronic delay built into the ICCD to compensate for the time difference between the two arms. Considering the polarization degeneracy effect of single-mode fibers, we placed a QHP combination at the exit end of the fiber to correct the outgoing polarization state. When the flip-flop mirror was upturned into the optical path, the outgoing signal photons were guided by the mirror to the SPAD to test the quality of the quantum state after passing through the QHP combination, and the QWP and HWP were adjusted by observing the conformal counts under different polarization basis vectors until the polarization state was completely corrected.

After obtaining a high quality quantum state, the reflector is lowered and the signal photons will lead to the imaging optical path. The signal light with a spot diameter of about 1 mm passes through the beam reduction system and then illuminates the phase object (PO) to be imaged, which is then amplified by an objective lens with a focal length of 20 mm and a 100 mm lens, resulting in a total system amplification of 7.5 times. After passing through the phase object, the signal light again passes through a 4F system with a metasurface placed at the center of the Fourier plane of the 4F system to modulate the spatial spectrum and photon polarization states, and finally the signal light enters the ICCD for imaging.

*3.2 Characterization of quantum light sources*

The brightness and fidelity of the polarization-entangled light source are key influences on image quality, so we first tested the quality of the light source before imaging. At a pump power of 73 mW, the single-photon count rates at the heralding and signal arm are 110 k/s and 210 k/s, respectively. The coincidence count (C) reaches a maximum of 250 k/s when the polarization choices of the heralding and signal arm are H, V or V, H, and the conformal count reaches a minimum of 0.6 k/s when the polarization choices are H, H or V, V. The contrast is $V = \frac{C_{max}-C_{min}}{C_{max}+C_{min}}$=95.3%, breaking the limit of Bell's Inequality. In addition, the effective prediction rate is measured to be 22.73.

However, the high contrast in the horizontal and vertical orthogonal polarization bases does not indicate the quality of the source, which could well be in other mixed states. Therefore, we measured the polarization interference curves and reconstructed the quantum state density matrices to show the entanglement properties of the two-photon. We fixed the HWP angle in the signal arm at 0° or 22.5°, while rotating the HWP angle of the trailer arm in 10° steps to obtain the polarization interference curve. As shown in Fig. 3(b), the measured coincidence counts in 1 s fit the sine function curve well. In addition, we performed quantum state tomograph of the polarization entangled source, and Fig. 3(c) shows the real and imaginary parts of the reconstructed density matrix, respectively. According to the definition of the fidelity: $F = \langle \Psi|\hat{\rho}|\Psi\rangle$, where $\Psi$ is the target quantum state and $\hat{\rho}$ is the density matrix reconstructed from the experimental data. The fidelity is calculated to be as high as 95.1% for the source in the pure state $1/\sqrt{2}(|HV\rangle + |VH\rangle)$.

*3.3 Higher signal noise ratio*

Based on the successful preparation of a high fidelity polarization-entangled light source and the systematic elucidation of the electromagnetic control mechanism and imaging principle of the metasurface, this study aims to demonstrate the superiority of the quantum imaging technology integrated with the metasurface device over the traditional imaging schemes. In particular, by exploiting the entanglement properties of photon pairs and the two-photon conformal measurement technique, quantum imaging is able to significantly improve the imaging contrast and signal noise ratio (SNR) of the system under the condition of maintaining the same number of incident photons, providing a new solution to break the technological bottleneck of traditional optical imaging.

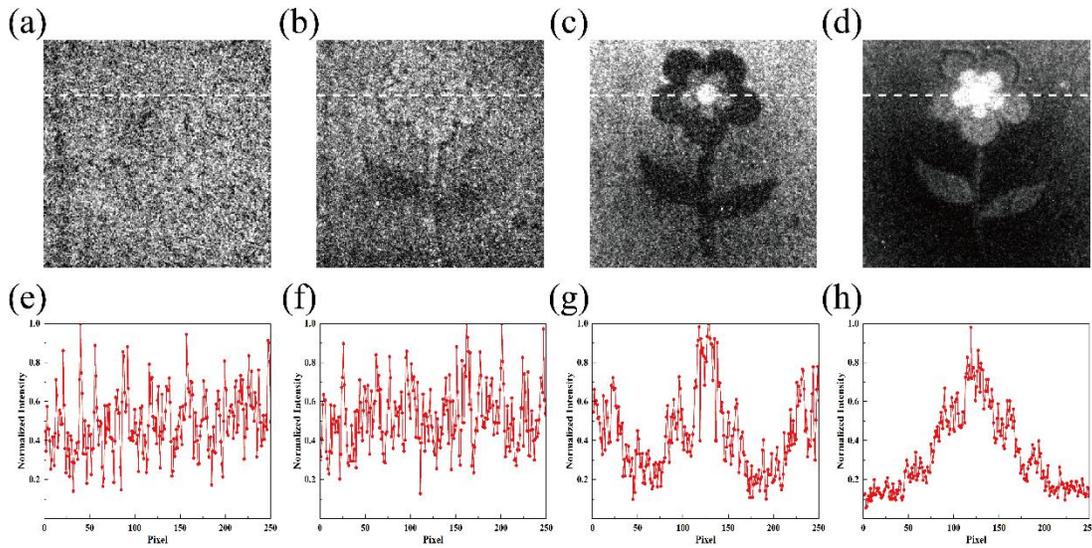

**Fig. 4 Experimental results of quantum bright and dark phase contrast imaging.** (a) Classic imaging; (b) Quantum imaging without metasurface modulation; (c) Quantum bright phase contrast imaging; (d) Quantum dark phase contrast imaging; (e-h) Intensity maps at the white dashed line.

To demonstrate the advantages of high contrast and high signal noise ratio of quantum imaging, we acquired images in classical mode and quantum imaging without metasurface modulation for comparison. Fig. 4(a) shows a classic image of a continuous collection mode (FO mode) triggered internally by the ICCD, where the camera gate is kept open during the acquisition and all signal photons of are captured during the acquisition time, including a large number of unassociated ambient noise photons that are continuously captured by the camera to the extent that they do not reveal obvious image features. However, the entangled photon pairs generated by utilizing spontaneous parametric down-conversion have strong correlation properties in the time domain and can be imaged by secondary correlation through the externally triggered mode (DDG mode) of the ICCD. When the SPAD detects a heralding photon, it triggers the time gate of the ICCD at the signal side to open briefly. In the experiment, we set the acquisition time to 20 min and the time width of the gate to 5 ns, so that only twin signal photons captured in the time gate during the acquisition time accumulate to form an image, thus significantly filtering out noise and realizing high-contrast and high-SNR imaging. Fig. 4(b), (c), and (d) show a bright phase contrast image and a dark phase contrast image without metasurface modulation, respectively, acquired by the DDG mode of the ICCD under the illumination of the quantum light source. Fig. 4(e-h) show the intensity maps of the light field varying along the dashed line in Fig. 4(a-d). Comparing the classical imaging effect of FO modes, it can be seen that even the DDG modes without metasurface modulation have shielded a lot of noise and shown imaging effect, but the contrast is very low. Only with the modulation of the metasurface can we realize the high-contrast bright and dark phase contrast imaging effect, so the light field modulation ability of the metasurface plays a key role in the imaging scheme. Based on the definition of contrast ratio $V = \frac{C_{max} - C_{min}}{C_{max} + C_{min}}$, the imaging contrast ratios of Fig. 4(a-d) are calculated to be 0.28, 0.36, 0.75, and 0.81, respectively. Therefore, quantum imaging has the advantage of higher SNR under the same imaging conditions.

## 3.4 Remote switching of imaging modes

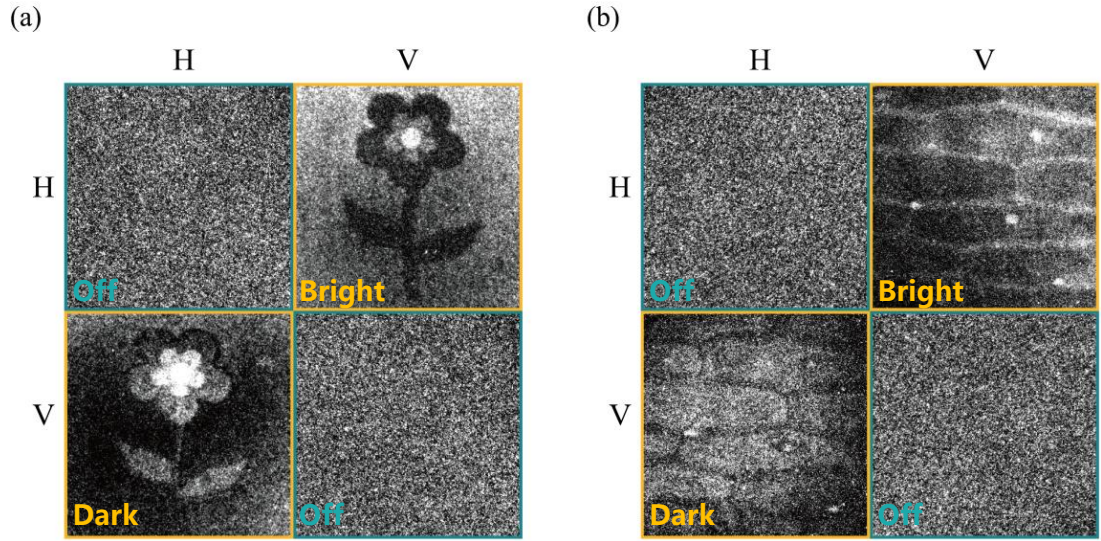

**Fig. 5 Remote switching of quantum bright-dark phase contrast imaging functions.** (a) Bright-dark phase contrast images of a low phase gradient sample. The number of acquisition frames is 1200, and the acquisition time of each frame is 1 s. (b) Quantum bright-dark phase contrast image of onion epidermis. The number of frames acquired is 3600, and the acquisition duration of each frame is 1 s. The horizontal axis represents the projected polarization state of the QHP in the signal arm, the vertical axis represents the projected polarization state of the QHP in the preview arm, and the images displayed in the coordinate system represent the images projected in different polarization states acquired by the ICCD camera and processed through noise.

Quantum imaging not only has the advantage of high SNR as described in the previous section, but also can utilize the polarization entanglement property of entangled photon pairs to realize the function of remotely switching the imaging modes of the imaging system, which improves the flexibility of this system. As shown in Fig. 5, these are the images obtained by projecting the signal and heralding arms on different polarization basis, where the horizontal (vertical) axis indicates the projected polarization state of the QHP in the signal (heralding) arm, and 1200 frames were acquired for each image, with an acquisition duration of 1 s per frame. The image in the orange border indicates that the projected polarization states of the signal arm and the heralding arm are orthogonally polarized bases when $1/\sqrt{2}(|HV\rangle + |VH\rangle)$ is satisfied to realize the bright-dark phase contrast image. And the green border indicates that the bright-dark phase contrast image cannot be realized when the projected polarization states of the signal arm and the heralding arm are the same. Fig. 5(a) demonstrates remotely switched quantum bright-dark phase contrast imaging of the low phase gradient sample, in which each image was acquired at 1200 frames with a 1-s acquisition duration per frame.

In biomedical imaging, conventional bright field microscopes are challenged by the low contrast and homogenization of transparent biological samples, while conventional staining methods may affect cellular activity, and this system proposes a new solution to this dilemma. Fig. 5(b) demonstrates quantum bright and dark phase contrast imaging of onion epidermal cells. When the polarization basis vectors of the two paths are projected to $|HV\rangle$ or $|VH\rangle$, the ICCD can acquire two high-contrast images of onion epidermal cells in either bright or dark phase contrast, and the switching of the two modes only needs to be executed at the trailer end, which realizes the remote manipulation function of image acquisition. The experimental results show that quantum imaging combined with metasurface modulation is more effective in imaging the onion epidermis. The technique exploits the non-classical properties of quantum-entangled light sources to enable high-contrast imaging under very low light conditions while avoiding phototoxicity to cell samples. This makes the introduction of quantum bright-dark phase contrast imaging techniques particularly important. Quantum bright-dark phase contrast imaging not only provides higher phase resolution and sensitivity, but also has the potential to reveal the dynamic processes of biological samples on a microscopic scale in the future, providing a more powerful tool for biomedical research, and showing great potential in the future in the fields of bio-imaging and life sciences.

## 4　Discussion

In conclusion, we have experimentally demonstrated that the combination of a quantum polarization entanglement source and a polarization-multiplexed metasurface in an optical imaging system can achieve high contrast imaging under low flux illumination for any combination of phases with remotely switchable bright-dark phase contrast imaging. We ultilize a ICCD camera to collect twin photons and accumulate them for imaging by using a quantum polarization entanglement source as a remote switching trigger, allowing remote switching of quantum bright-dark phase contrast imaging modes without modifying the imaging optical paths of the sample and metasurface in the imaging system. The system not only achieves high signal noise ratio imaging of low phase gradient samples, but also demonstrates flexibility and high integration and shows new application potentials and broad prospects for quantum communication, information security, and other fields. In addition, the experimental results show that the system is also applicable to the field of biomedical imaging. Quantum bright-dark phase contrast imaging technology can further enhance the imaging contrast and resolution of biological samples, with the advantages of no labeling and no damage, which is an important development in the field of biomedical imaging, and has a wide range of future applications in the fields of life sciences and biological research.

**Caption List**

**Fig. 1** Schematic representation of a switchable quantum bright-dark phase contrast imaging system.

**Fig. 2** Characterization of experimental samples.

**Fig. 3** Experimental setup and characterization of entangled light source.

**Fig. 4** Experimental results of quantum bright and dark phase contrast imaging.

**Fig. 5** Remote switching of quantum bright-dark phase contrast imaging functions.